\documentclass{vow2008}

\title{Scale Length of Disk Galaxies}
\author{Kambiz Fathi}
\affil{Department of Astronomy, Stockholm University, 106 91 Stockholm, Sweden}
\author{Mark Allen}
\affil{Observatoire de Strasbourg, UMR 7550 Strasbourg 67000, France}
\author{Eduardo Gonzalez-Solares}
\affil{Institute of Astronomy, University of Cambridge, Madingley Road, Cambridge CB3 0HA, UK}
\author{Evanthia Hatziminaoglou}
\affil{European Southern Observatory, Karl-Schwarzschild-Str. 2, 85748 Garching bei M\"unchen, Germany}
\author{Reynier Peletier}
\affil{Kapteyn Astronomical Institute, Postbus 800, 9700 AV Groningen, The Netherlands}

\usepackage{graphicx}
\begin{document}


\maketitle

\begin{abstract}
As a part of a Euro-VO research initiative, we have undertaken a programme aimed at studying the scale length of 54909 Sa-Sd spiral galaxies from the SDSS DR6 catalogue. We have retrieved $u, g, r, i, z$-band images for all galaxies in order to derive the light profiles. We also calculate asymmetry parameters to select non-disturbed disks for which we will derive exponential disk scale lengths. As images in different bands probe different optical depths and stellar populations, it is likely that a derived scale length value should depend on waveband, and our goal is to use the scale length variations with band pass, inclination, galaxy type, redshift, and surface brightness, in order to better understand the nature of spiral galaxies.
\end{abstract}

\section{Introduction}

The exponential scale length of a galaxy disk is one of the most fundamental parameters to determine its morphological structure, as well as to model its dynamics, and the fact that the light distributions are exponential makes it possible to contrain the formation mechanisms (Freeman 1970). The scale length determines how the stars are distributed throughout a disk, and can be used to derive its mass distribution, assuming a specific M/L ratio. Ultimately, this mass distribution is the primary constraint for determining the formation scenario (e.g., Dutton 2008, and references therein), which dictates the galaxy's evolution. As the galaxy evolves substructures such as bulges, pseudo-bulges, bars, rings, and spiral arms may build up, which in turn considerably change the morphology of the host disks (e.g., Combes \& Elmegreen 1993). Analytic disk formation scenarios (e.g., Lin \& Pringle 1987) predict that in cases where angular momentum is conserved, the disk scale length is determined by the the angular momentum profile of the initial cloud, and the scale length in a viscous disk is set by the interplay between star formation and dynamical friction (e.g., Silk 2001). These processes form the basis of a galaxy's gravitational potential and the strength of gravitational perturbations, the location of resonances in the disk, the formation and evolution of spiral arms and bars, and the dynamical feeding of circumnuclear starbursts and nuclear activity (e.g., Elmegreen et al. 1996; Fathi et al. 2008).

Photometrically, the scale length is derived by azimuthally averaging profiles of the surface brightness which is in turn decomposed into a central bulge and an exponential disk, accounting for other components such as bars and rings. 

As images in different bands probe different optical depths and stellar populations, it is likely that a derived scale length value should depend on waveband. Dusty disks are more opaque and often deliver larger scale length values in bluer bands when compared with red and/or infrared images. Similar effects can also be caused by the stellar populations. These observational effects thus not only give us insights about the disks that we are studying, but also need to be quantified for a better comparison between different data sets and galaxy types. Both the effects of stellar populations and dist extinction have been subject to much discussion over the years (e.g., Simien \& de Vaucouleurs 1983; Valentijn 1990; van Driel et al. 1995; Peletier et al. 1995; Beckman et al. 1996; Prieto et al 2001; Graham \& de Blok 2001; MacArthur 2003; Cunow 1998, 2001, 2004). A detailed and extensive analysis of the dust effects has also been presented for a few tens of galaxies in Holwerda (2005) and subsequent papers, however, as noted by Peletier et al. (1994) and van Driel et al. (1995), the scale length alone in different band passes in small sample cannot be used to break the age/metallicity and dust effects. Investigating the scale length variation as a function of inclination for large numbers of galaxies is necessary to distinguish between the dust and population effects.

The common denominator in all the previous studies is the roughly comparable sample sizes. Most studies have so far analysed individual galaxies, or samples containing a few tens of galaxies (see table 1). This is not to be mistaken with the number of great results from the Sloan Digital Sky Survey (SDSS) studies in the last years, but these works have not studied the astrophysical effects mentioned here. We have undertaken a programme that aims at quantifying how the disk scale length varies with band pass, inclination, galaxy type, redshift, and surface brightness. We have searched the entire SDSS Data Release 6 (DR6) data set and have selected 54909 spiral galaxies suitable for our analysis. Here, we present a description of our study along with some preliminary results.

\begin{table}
  \begin{center}
    \caption{Sample sizes for the work mentioned in the text (in alphabetic order).}
    \renewcommand{\arraystretch}{1.2}
    \begin{tabular}[h]{lr}
      \hline
      Reference			& Number of galaxies \\
      \hline\hline
      Cunow (1998, 2001, 2004)	& 14, 60, 39 \\
      Graham \& de Blok (2001)	& 120 \\
      MacArthur et al. (2003)		& 121\\
      Peletier et al. (1995)		& 37 \\
      Prieto et al. (1996)		& 15 \\
      Simien \& de Vaucouleurs (1983)	& 98 \\
      van Driel et al. (1995)		& 55 \\
      \hline \\
      \end{tabular}
    \label{tab:table}
  \end{center}
\end{table}

\begin{figure}
\begin{center}
 \includegraphics[width=.49\textwidth]{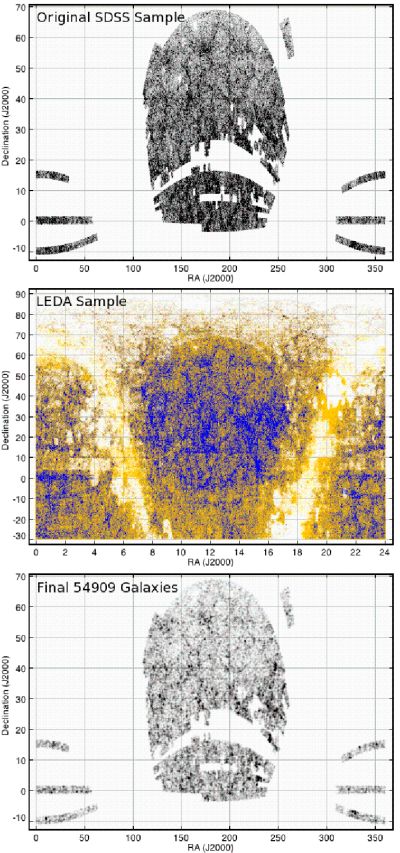}
  \caption{Top: Right ascension and declination distribution of the SDSS sample fulfilling the first sorting criteria (i.e., galaxies with spectroscopic redshift, larger than 10 pixels, extinction smaller than $A_V=1.0$, and with inclination between $15^\circ$ and $70^\circ$), Middle: The total LEDA catalogue for which the LEDA services provide a hubble classification number (0=S0a, 1=Sa, 2=Sab, etc.), and Bottom: Our final sample fulfilling all our selection criteria described in the text.} 
  \label{fig:design}
\end{center}
\end{figure}

\begin{figure*}
\begin{center}
\includegraphics[width=.49\textwidth]{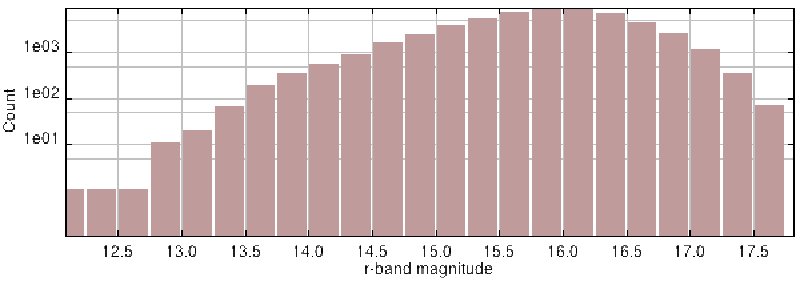}
\includegraphics[width=.49\textwidth]{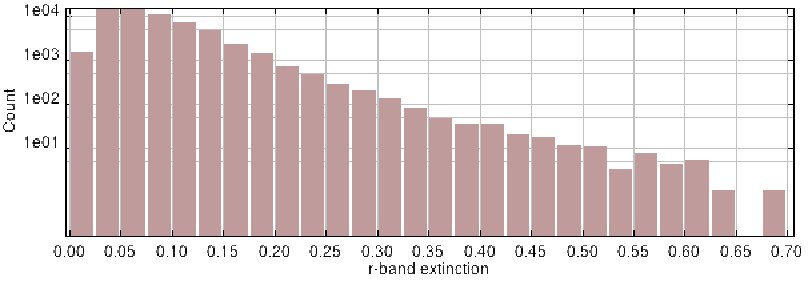}
\includegraphics[width=.49\textwidth]{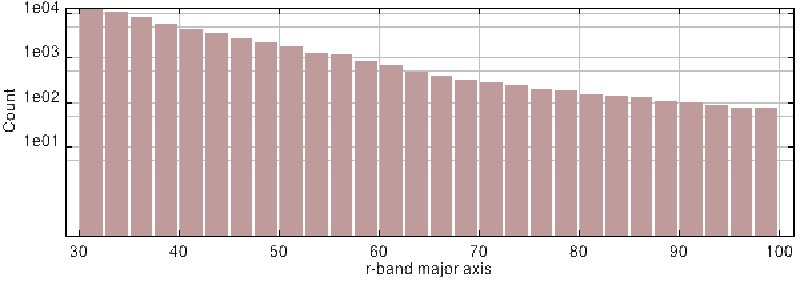}
\includegraphics[width=.49\textwidth]{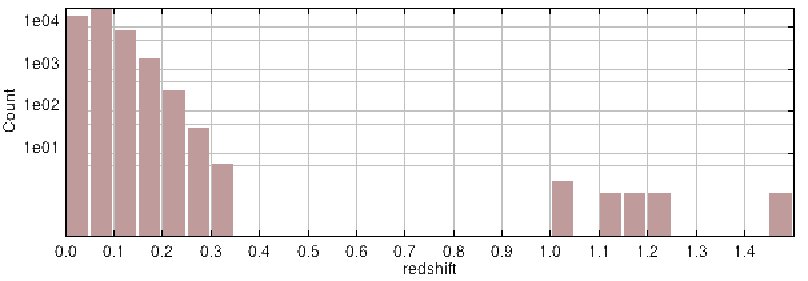}
\includegraphics[width=.49\textwidth]{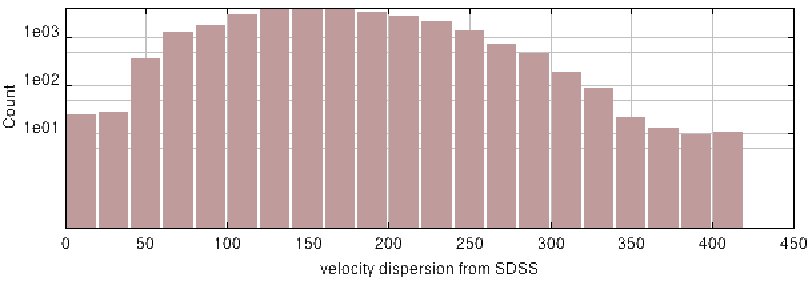}
\includegraphics[width=.49\textwidth]{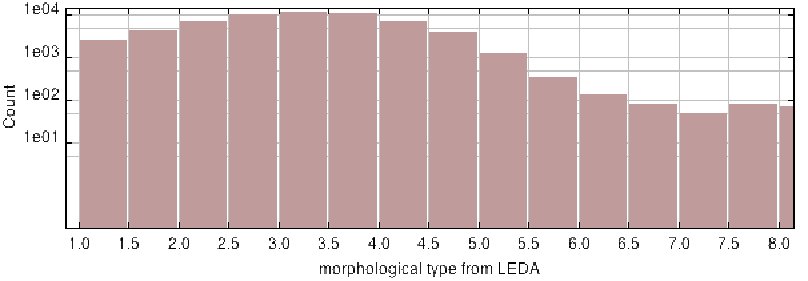}
  \caption{Distribution of some key parameters (magnitude, Galactic extinction, major axis, redshift, velocity dispersion, and morphological type) retrieved from the SDSS and LEDA database.} 
  \label{fig:design}
\end{center}
\end{figure*}

\section{Sample selection}
The sample was selected by searching the SDSS DR6 catalogue and cross matching with the LEDA\footnote{http://leda.univ-lyon1.fr/} catalogue (Paturel et al. 2003) to retrieve Hubble classifications. As the SDSS provides a number of morphological as well as kinematic parameters, which we use to constrain our sample against biases. Our first requirements ensure that for each galaxy we have reliable redshift measurement, low extinction, and number of pixels sufficient to derive a light profile with a good coverage of the disk region. Moreover, we decided to ensure that our galaxies do not contain edge-on or face-on galaxies to avoid selection effect problems. We first retrieved tabular data for all SDSS galaxies for which excellent image quality is delivered, are larger than 30 pixels, spectroscopic redshifts are available, have extinction $A_V \le 1.0$ mag, have inclination $15^\circ < i < 70^\circ$. We retrieved a total of 475408 galaxies, and first investigated the smallest numbers of pixels needed to resolve the disk. We found that a minimum of 70 pixels are needed, thus removed all galaxies with major axis (in $r$-band) smaller than 70 pixels (28 arcsec) to ensure that the images cover the disk region.

We made use of LEDA by first retrieving the entire catalogue. As this service provides a numeric Hubble classification parameter, we could easily select all the spiral galaxies, which we later cross-correlated with the SDSS catalogue. We found a total of 54909 Sa-Sd spiral galaxies (see Fig.~1), for which all the morphological and spectroscopic parameters from SDSS and LEDA were stored, and $u, g, r, i, z$-band images were to be downloaded. It should be noted that at this stage, we are unable to determine whether the galaxies in our sample are isolated or disturbed systems, as this information is not provided by any of the catalogues we have used. We make this distinction using the asymmetry parameter described in Conselice (2003).

\begin{figure}
\begin{center}
 \includegraphics[width=.49\textwidth]{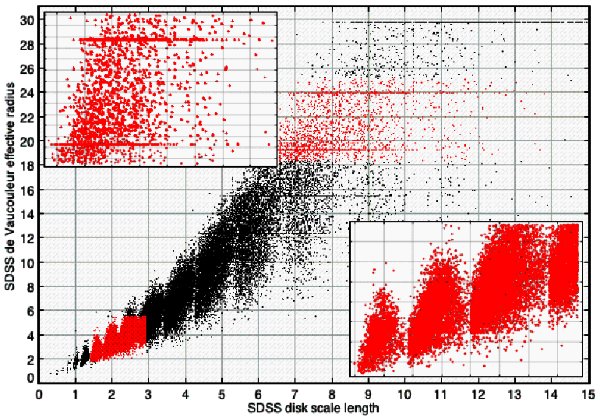}
  \caption{De Vaucouleurs effective radius ($y$-axis) versus disk scale length ($x$-axis) provided by the SDSS team for all 54909 galaxies in our final sample. Although these numbers are good first guesses, the odd clustering of the data points lead us the to conclusion that we re-derive the scale lengths. The insets illustrate the strange clustering of the points (coloured data points) provided by the SDSS.} 
\end{center}
\end{figure}

The first question that rises at this point is the fact that SDSS delivers the disk scale length as well as de Vaucoulers effective radius for each galaxy (in all bands), and that these values could be used to carry out or analysis. In Fig.~2, we show that the values provided by the SDSS team show anomalies that are beyond our satisfaction for carrying out our analysis. The plot shows a strange "clustering" of the effective radii and scale lengths around some numbers, the source for which we cannot find. We thus decide to re-calculate the scale lengths.

\begin{figure*}
\begin{center}
 \includegraphics[width=.75\textwidth]{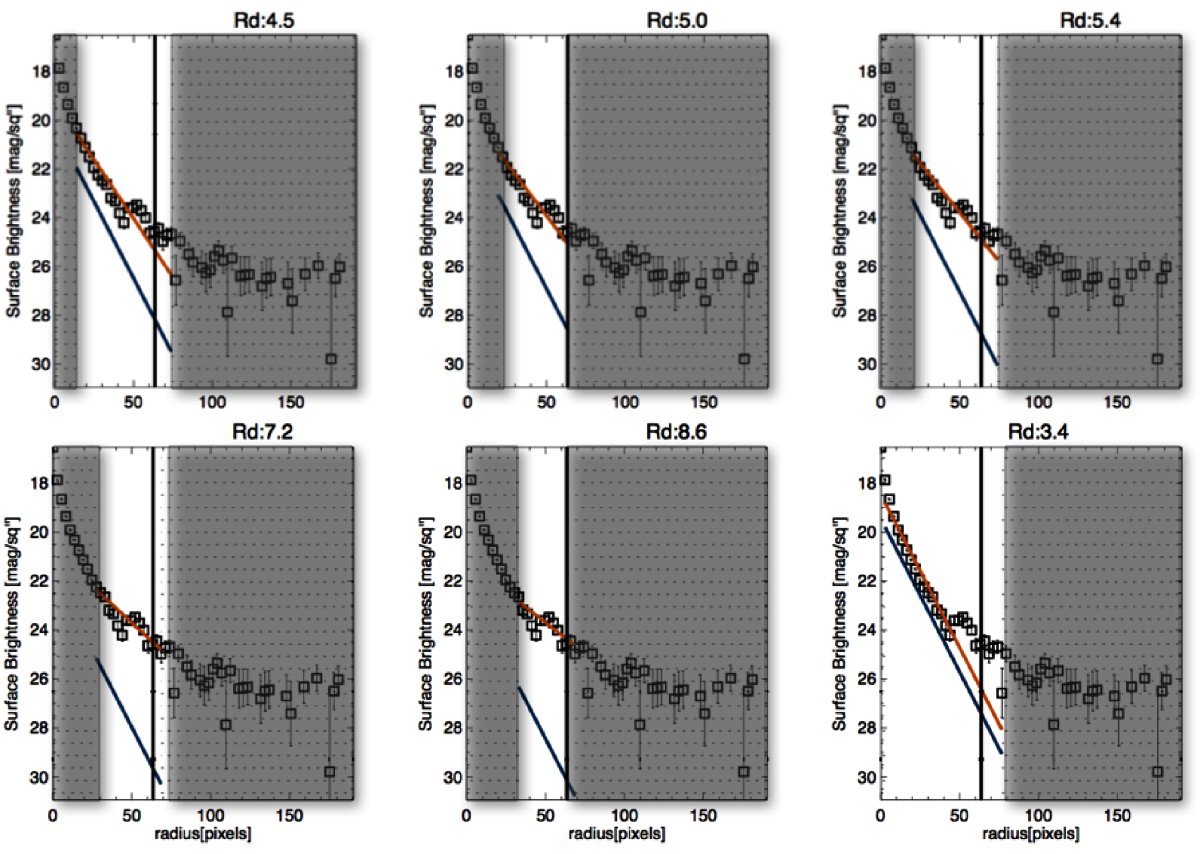}
  \caption{Illustrating the fits to the different regions of the light profile. Fitting different regions results in different scale length values, and as shown in the bottom right panel, fitting over the entire galaxy reproduces the value provided by the SDSS, which is illustrated by the "blue" line (arbitrarily shifted (in $y$-axis). Printed at the top of each panel, is the corresponding scale length value.} 
\end{center}
\end{figure*}

Various Virtual Observatory (VO) methods were investigated to perform the download of the SDSS images. The SkyView \footnote{http://skyview.gsfc.nasa.gov/} was chosen for this task. This service has the advantage of being able to create fits files centred at a given sky coordinate and with a pre-specified size. The image size is an important parameter for achieving a reliable an accurate sky subtraction, thus we require that the images are $900 \times 900$ pixels for the sky region to be sampled for all galaxies. Moreover, SkyView is able to re-scale the image backgrounds to the same level, hence correcting for differences between the SDSS plates. We have experienced that a linear image download could be a tedious process. With an image size of 3.2 MB/image, and at a constant rate of 0.2 MB/s, the download would require at the very best and with a continuous connection $\approx 51$ days. We have therefore carried out the download using multiple parallel data requests to the SkyView service using the python scripting language, and utilising 3 individual computers. This allowed us to download the full sample over approximately 13 days.

\begin{figure}
\begin{center}
 \includegraphics[width=.49\textwidth]{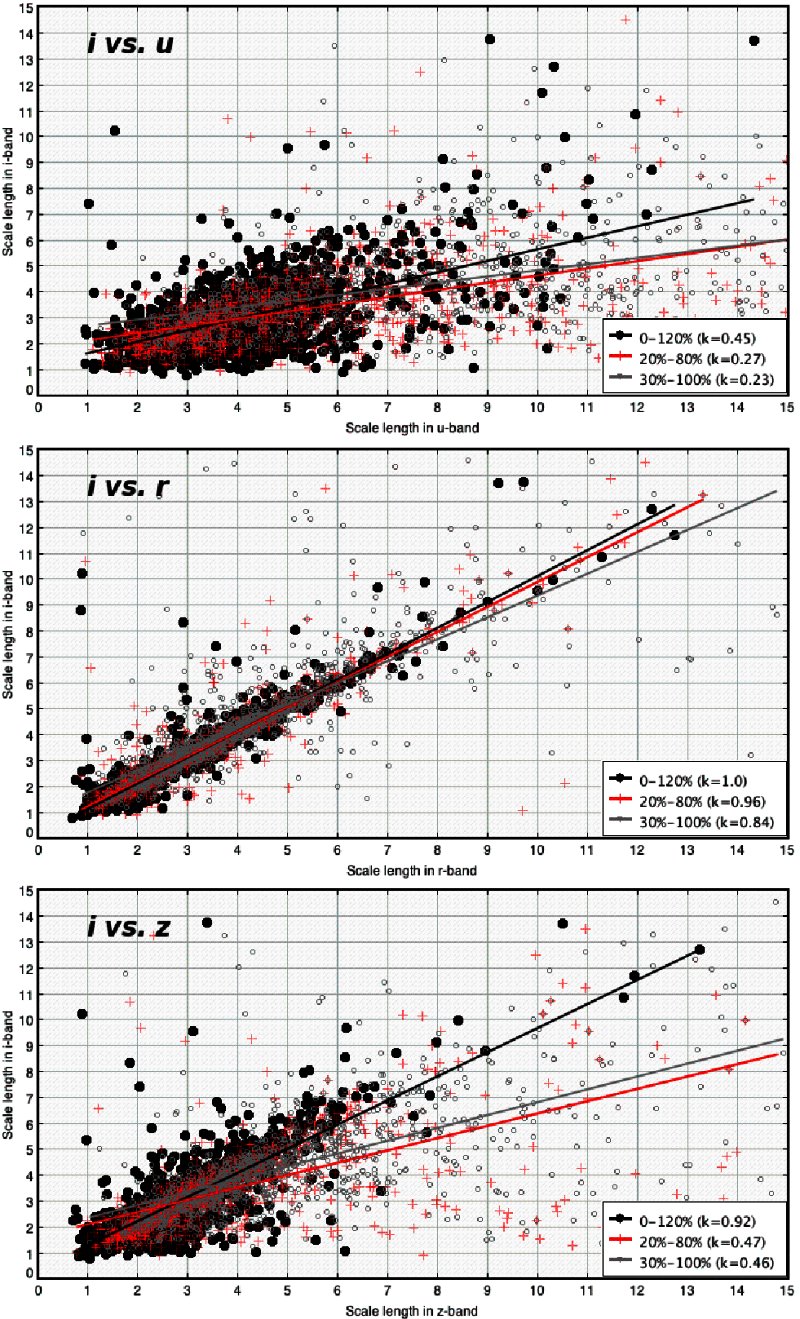}
  \caption{Scale lengths for a sub sample of 1315 randomly selected galaxies in $u$, $r$, $i$, and $z$-band. Here, fits to three different disk regions are plotted to show that fitting some outer region is not necessarily the best strategy. The relatively large scatter in the top and bottom panels also demonstrate that the images are not very deep in $u$ and $z$ bands. At the bottom right corner of panel, the fitted regions are printed, where 0-120\% means that the light profile between galactocentric radius of $r=0*iso_A$ and $r=1.2*iso_A$ was fitted by one single exponential profile, etc. .} 
\end{center}
\end{figure}

\section{Deriving the light profile}
We derive the disk scale length using standard IDL routines and make use some important parameters included in the SDSS information in order to constrain galaxy geometry as well as the location of the sky region. These are $iso_B$, $iso_A$, $iso_{Phi}$, and for consistency, we use these quantities in $r$-band. The procedure carries out the following steps:
\begin{itemize}
\item Reading the image, and calculating the asymmetry parameter $A=\sum (I-I_{180})/\sum I $, where $I$ is the galaxy image and $I_{180}$ is the image rotated by 180 degrees around the galaxy centre. 
\item Selecting the sky region as the ellipse encompassing the region between $2.0*iso_A$ and $2.5*iso_A$. The mean value of this region, using Tukey's bi-weight mean formalism described in Mosteller \& Tukey (1977), is used for the sky subtraction as well as setting the background level.

\item To remove nuisance stars and point sources from the image, we extract point sources with SExtractor (Bertin \& Arnauts 1996), but storing pixels belonging to all point sources that are larger than 4 pixels and more than 3 $\sigma$ above the background level. All pixels belonging to these sources are then masked out.

\item Using the $iso_B$, $iso_A$, $iso_{Phi}$ parameters from SDSS, we then section each galaxy into 2-pixels wide ellipses oriented at $iso_{Phi}$ and with minor-to-major axis ratio $b/a=iso_B/iso_A$. The mean surface brightness value within each ellipse is calculated, to compile the galactocentric light profile for each galaxy image.

\item As experience has shown that light profile decomposition in a disk together with any other component (e.g., Fathi et al. 2003) introduces complications that are not necessary for the nature of our analysis. We thus derive the disk scale length simply by fitting an exponential profile to the disk region of each galaxy. We determine the region of interest by empirically fitting eight different regions covering the range between 15\%-30\% and  85\%-115\% of the $iso_A$ parameter. This procedure means that we are simply cutting out the central regions of the galaxies where bulges and strong bars are expected. Fig.~4 shows the result of such a test where we find that (in many cases) we are able to derive a scale length comparable to the value from the SDSS catalogue. It should be noted that this plot is only for one galaxy, and for regions outside what is noted here, and we have tested more regions than the eight regions detailed above here.

\end{itemize}
Assuming that the $r$ and $i$-band images probe similar stellar populations and dust content, we are able to use the correlation of the scale length values between these bands to find the optimal region for deriving the scale lengths. Fig.~5, illustrates for a preliminary and randomly selected sub-sample of 1315 galaxies, that using a small outer region is not the best way to derive the scale length, but for this sub-sample is seems that fitting the galaxy light profile over $1.2*iso_A$ gives the best result. We plan to carry out this test for the entire sample.

\section{Current status}
We have currently downloaded the $u,g,r,i,z$ images for all 54909 galaxies, and we are in the process of calculating the scale length using different regions from each image. With the data at hand, we will be able to first remove highly asymmetric galaxies to minimise the use of disturbed disks. We will then derive the disk scale lengths for all the "isolated" disk galaxies, and statistically explore how this parameter changes as a function of inclination, band, redshift, etc. Moreover, we plan to cross-correlate our sample with the Two Micron All Sky Survey to further explore the scale lengths also in $J, H, K$ bands and to further explore the dust and stellar population effects.

\section*{Acknowledgments}
We thank the organisers, LOC, and STOC, for creating this stimulating opportunity to exchange ideas between different disciplines. This work makes use of EURO-VO software, tools or services. The EURO-VO has been funded by the European Commission through contract numbers RI031675 (DCA) and 011892 (VO-TECH) under the 6th Framework Programme and contract number 212104 (AIDA) under the 7th Framework Programme. We also acknowledge the use of NASA's SkyView facility (http://skyview.gsfc.nasa.gov) located at NASA Goddard Space Flight Center, the usage of the HyperLeda database (http://leda.univ-lyon1.fr), and the TOCAT software (http://www.starlink.ac.uk/topcat/).



\begin{thebibliography}{}
\bibitem[Aguerri et al (1998)]{Aetal98} Aguerri, J. A. L. et al. 1998, AJ, 116, 2136
\bibitem[Bertin \& Arnouts (1996)]{BA96} Bertin, E. \& Arnouts, S., 1996, A \& AS, 117,393
\bibitem[Beckman et al. (1996)]{Betal96} Beckman, J. E. et al. 1996, ApJ, 467, 175
\bibitem[Bertin \& Arnouts (1996)]{BA96} Bertin, E. \& Arnouts, S. 1996, A\&AS, 117, 393
\bibitem[Combes \& Elmegreen (1993)]{CE93} Combes, F., \& Elmegreen, B. G. 1993, A\&A, 271, 391
\bibitem[Conselice (2003)]{Conselice03} Conselice, C. 2003, ApJSS, 147, 1
\bibitem[Cunow (1998)]{Cunow98} Cunow, B. 1998, A\&ASS, 129, 593
\bibitem[Cunow (2001)]{Cunow01} Cunow, B. 2001, MNRAS, 323, 130
\bibitem[Cunow (2004)]{Cunow04} Cunow, B. 2004, MNRAS, 353, 477
\bibitem[Dutton (2008)]{Dutton08} Dutton, A. 2008, MNRAS, subm. (arXiv:0810.5164) 
\bibitem[Elmegreen et al. (1996)]{Eetal96} Elmegreen, B. G. et al. 1996, AJ, 111, 2233  
\bibitem[Fathi \& Peletier (2003)]{FP03} Fathi, K. \& Peletier, R. F. 2003, A\&A, 407, 61 
\bibitem[Fathi et al. (2008)]{Fetal08} Fathi, K. et al. 2008, ApJ, 675, L17 
\bibitem[Freeman (1970)]{Fetal08} Freeman, K. C. 1970, 160, 811
\bibitem[Graham \& de Blok (2001)]{GdeB01} Graham, A. W., de Blok, W. J. G. 2001, ApJ, 556, 177
\bibitem[Holwerda (2005)]{Holwerda05} Holwerda, B. 2005, PhD. thesis, Univ. of Groningen
\bibitem[Simien \& de Vaucouleurs (1983)]{SdeV83} Simien, F. \& de Vaucouleurs, G. 1983, IAUS, 100, 375
\bibitem[Lin \& Pringle (1987)]{LP87} Lin, D. N. C. \& Pringle, J. E. 1987, MNRAS, 225, 607 
\bibitem[MacArthur (2003)]{MacA03} MacArthur, L. A., Courteau, S., Holtzman, J. A. 2003, ApJ, 582, 689
\bibitem[Mosteller \& Tukey (1977)]{MT77} Mosteller, F., Tukey, J. 1977, Data Analysis and Regression, Addison-Wesley
\bibitem[Paturel et al. (2003)]{Petal03} Paturel G. et al. 2003, A\&A, 412, 45
\bibitem[Peletier et al. (1994)]{Petal94} Peletier, R. F. et al. 1994, A\&AS, 108, 621
\bibitem[Peletier et al. (1995)]{Petal95} Peletier, R. F. et al. 1995, A\&A, 300, L1
\bibitem[Prieto et al. (2001)]{Petal01} Prieto, M. et al. 2001, A\&A, 367, 405
\bibitem[Silk (2001)]{Silk01} Silk, J. 2001, MNRAS, 324, 313 
\bibitem[Valentijn (1990)]{Val90} Valentijn, E. A. 1990, Nature, 346, 153
\bibitem[van Driel et al. (1995)]{vDetal95} van Driel, W. et al. 1995, A\&A, 298, 41 
\end{thebibliography}
\end{document}